\documentclass[journal=langd5,manuscript=article]{achemso}

\usepackage[version=3]{mhchem} 
\usepackage{amsmath}
\usepackage{amssymb}
\usepackage{subcaption}
\usepackage{graphicx}

\author{Niladri Sekhar Mandal}
\author{Ayusman Sen}
\affiliation[Pennsylvania State University]
{Department of Chemical Engineering, Pennsylvania State University, University Park, PA 16802}
\alsoaffiliation[Pennsylvania State University]
{Department of Chemistry, Pennsylvania State University, University Park, PA 16802}
\email{asen@psu.edu}

\title[An \textsf{achemso} demo]
  {The Relative Diffusivities of Bound and Unbound Protein Can Control Chemotactic Directionality}


\begin{document}

\begin{tocentry}
    \includegraphics{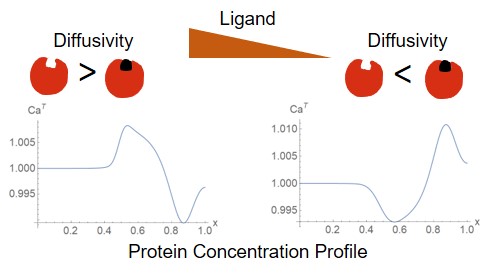}
\end{tocentry}

\begin{abstract}
Enzyme-based systems have been shown to undergo directional motion in response to their substrate gradient. Here, we formulate a kinetic model to analyze the directional movement of an ensemble of protein molecules in response to a gradient of the ligand. A similar analysis has been performed to probe the motion of enzyme molecules in response to a gradient of the substrate under catalytic conditions. In both cases, a net movement up the ligand/substrate gradient is predicted when the diffusivity of the ligand/substrate-bound protein is lower than that of the unbound protein (positive chemotaxis). Conversely, movement down the ligand/substrate gradient is expected when the diffusivity of the ligand/substrate-bound protein is higher than that of the unbound protein (negative chemotaxis). The work underscores the critical importance of measuring the diffusivity of the bound protein and compare it with that of the free protein.
\end{abstract}

\section{Introduction}
Chemotaxis is a phenomenon historically used to describe the motion of organisms like bacteria towards or away from a chemical attractant or repellant\cite{Berg1975}, respectively. Recent studies have shown that ensembles of single enzyme molecules\cite{Sengupta2013,Dey2014,XiZhao2018, Jee2018}, as well as enzyme-attached microparticles\cite{Somasundar2019, Dey2015, Ji2019, Joseph2017, Wang2020}, also chemotax in response to a gradient of the substrate. Several theories to this date have been used to describe chemotaxis in enzyme-based systems\cite{Feng2020}. These include space-dependent diffusivity, phoretic mechanisms\cite{Agudo-Canalejo2018}, thermodynamically induced drift\cite{Schurr2013}. Phoretic mechanisms have not found wide-range experimental support\cite{Sengupta2013,Zhao2018}. Chemotaxis arising from a thermodynamic force has typically been accounted for using a cross-diffusion coefficient\cite{Schurr2013, Mohajerani2018} which dictates the movement of the enzyme towards or away from the substrate. It has also been proposed that enzymes may have different diffusivities when free and substrate-bound\cite{Agudo-Canalejo2018}. In this paper, we analyze the phenomenon of chemotaxis from a kinetic standpoint. Using two scenarios, a) substrate binding only and b) catalytic turnover of the substrate, we show that the direction of chemotactic migration depends critically on the relative diffusivities of the free and substrate-bound protein (enzyme). Unlike the previously used Finite Elements Method, we use the Finite Difference Method which is more appropriate for a regular geometry to solve the reaction-diffusion equations\cite{Blazek2015}. Typically, equilibrium conditions\cite{Schurr2013} or Michael-Menten kinetics\cite{Mohajerani2018} has been used previously to derive the thermodynamic force in chemotaxis. We discuss the limits where equilibrium or Michaelis- Menten kinetics can be assumed, and where these assumptions might break down.

\section{Methods}
\begin{figure}
  \includegraphics[scale =0.5]{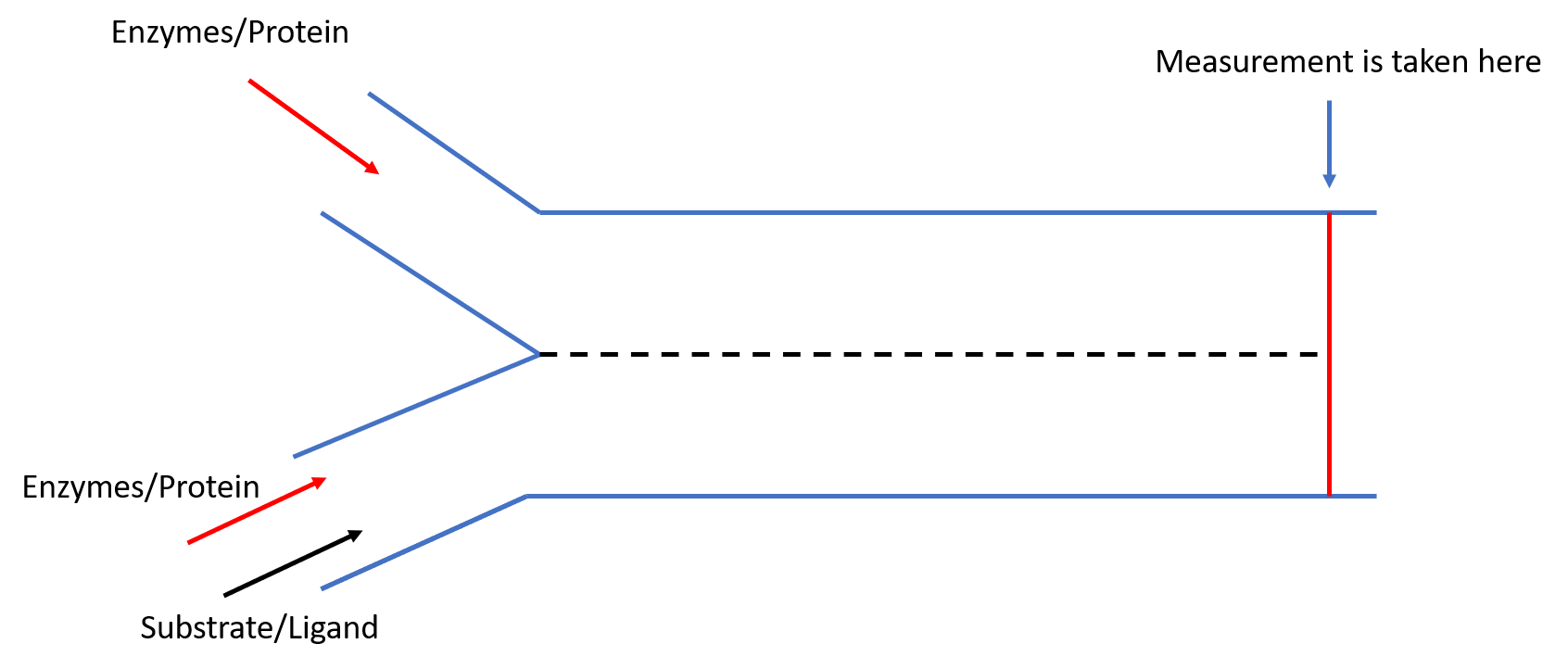}
  \caption{Microchannel used for chemotaxis experiments. The protein/enzyme is introduced from both the inflow channels. The ligand/substrate is introduced from one of the inflow channels. The velocity in the system is low enough to avoid intermixing of species at the beginning of the channel. The concentration profile of the enzyme is then measured at the outlet of the channel\cite{Sengupta2013}}
  \label{fgr:microchannel}
\end{figure}

\begin{figure}
  \includegraphics[scale =1]{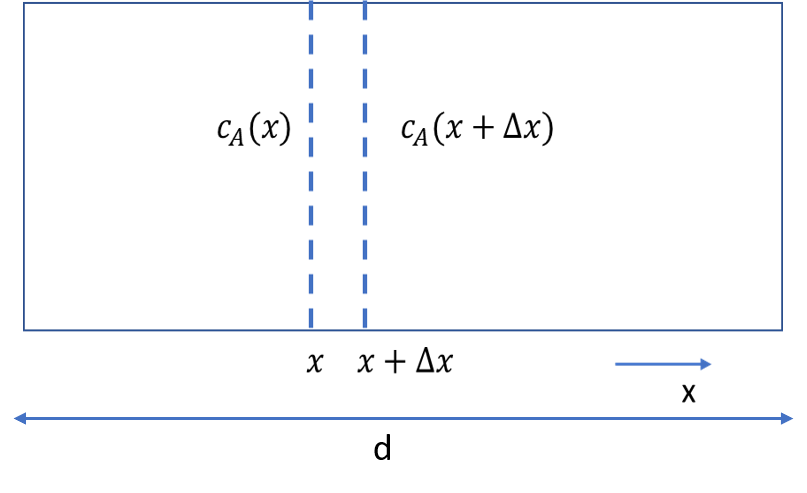}
  \caption{1-D system used for kinetic analysis in this work. The domain is one-dimensional and its width in the x-direction is d}
  \label{fgr:1Dmicro}
\end{figure}

In this section, we develop the relevant equations required to analyze the interaction between enzyme/protein with its substrate/ligand. For the rest of the paper, we refer to protein and ligand when we consider simple binding interactions. For catalytic reactions involving turnover, we use the terms enzyme and substrate. We first develop the equations for a binding interaction and then show how the set of equations changes when a catalytic step is added. We base our work on experiments of chemotaxis in microfluidic channels\cite{Lin2006, Diao2006,Sengupta2013,Dey2014,XiZhao2018} a simple diagram of which is shown in Figure~\ref{fgr:microchannel}. We begin our analysis of chemotaxis by assuming a one-dimensional system with no flow and channel width, $d$. This is shown in Figure~\ref{fgr:1Dmicro}.

\paragraph{Binding Interaction between a protein and a ligand}
Consider a binding reaction between a protein(A) and ligand(B) as follows:
\begin{equation}\label{eqn:rxn_main}
\ce{A + B <=>[k_1][k_{-1}] AB}
\end{equation}
The mass-transfer equations then can be written for all three species in the system, namely A, AB, and B \cite{Fogler2006}:
\begin{equation}\label{eq:mass_bal_A}
\frac{\partial c_A}{\partial t} = D_{A} \frac{\partial^2 c_A}{\partial x^2} - k_1 c_A c_B + k_{-1} c_{AB}
\end{equation}
\begin{equation}\label{eq:mass_bal_AB}
\frac{\partial c_{AB}}{\partial t} = D_{AB} \frac{\partial^2 c_{AB}}{\partial x^2} + k_1 c_A c_B - k_{-1} c_{AB}
\end{equation}
\begin{equation}\label{eq:mass_bal_B}
\frac{\partial c_B}{\partial t} = D_{B} \frac{\partial^2 c_B}{\partial x^2} -  k_1 c_A c_B + k_{-1} c_{AB}
\end{equation}
Here $c_i$ and $D_i$ stands for the concentration and the diffusion coefficient of the $i^{th}$ species respectively. We now want to investigate whether the assumption of equilibrium condition be universally valid for this system. The initial concentration of the protein and ligand in the system are $c_{A0}$ and $c_{B0}$ respectively. The concentration of the protein-ligand complex can be scaled with equilibrium assumption. The dissociation constant is given by $K_d = \frac{k_{-1}}{k_1}$. Then the variables can be scaled as follows\cite{Deen2011}:
\begin{equation}\label{eq:scaling_B}
    c_A \sim c_{A0} \hspace{1cm} c_B \sim c_{B0} \hspace{1cm} x \sim d \hspace{1cm} t \sim \frac{d^2}{D_A}  \hspace{1cm} c_{AB} = \frac{c_A c_B}{K_d}  
\end{equation}
Non-dimensionalizing Eq. ~\ref{eq:mass_bal_A} - ~\ref{eq:mass_bal_B} leads to the following equations:
\begin{equation}\label{eq:nondim_A}
\epsilon \frac{\partial \bar{c_A}}{\partial \bar{t}} = \epsilon \frac{\partial^2 \bar{c_A}}{\partial \bar{x}^2} - \bar{c_A} \bar{c_B} + \bar{c_{AB}}
\end{equation}
\begin{equation}\label{eq:nondim_AB}
\delta \frac{\partial \bar{c_{AB}}}{\partial \bar{t}} = \delta \gamma \frac{\partial^2 \bar{c_{AB}}}{\partial\bar{x}^2} + \bar{c_A} \bar{c_B} - \bar{c_{AB}}
\end{equation}
\begin{equation}\label{eq:nondim_B}
\eta \frac{\partial \bar{c_B}}{\partial \bar{t}} = \eta \beta \frac{\partial^2 \bar{c_B}}{\partial \bar{x}^2} - \bar{c_A} \bar{c_B} + \bar{c_{AB}}
\end{equation}
Non-dimensionalizing the mass-transfer equations give us the three Damkohler numbers for the system which are $\frac{d^2 k_1 C_{B0}}{D_A}, \frac{d^2 k_{-1}}{D_A}, \frac{d^2 k_1 C_{A0}}{D_A}$. The Damkohler numbers represent the ratio of time taken for diffusion to the time taken for reaction. For example, in Eq.~\ref{eq:nondim_A} the characteristic time for diffusion is $t_D = d^2/D_A$ and the characteristic time for reaction is $t_R =\frac{1}{k_1C_{B0}}$, hence the Damkolher number for Eq.~\ref{eq:nondim_A} is $\frac{d^2 k_1 C_{B0}}{D_A}$. The inverse of the Damkohler's numbers are characterized by the constants $\epsilon = \frac{D_A}{d^2 k_1 C_{B0}}$, $\delta = \frac{D_A}{d^2 k_{-1}}$ and $\eta = \frac{D_A}{d^2 k_1 C_{A0}}$. $\gamma = D_{AB}/D_A$ and $\beta = D_B/D_A$ are ratio of diffusivity of bound-protein and the free protein and the ligand and the free protein respectively. These constants appear in the mass-balance equations as evident from Eq.~\ref{eq:nondim_A}-Eq.~\ref{eq:nondim_B}. 
\\
At this point, we want to investigate whether we can assume equilibrium for the binding reaction in the domain. First, we introduce a notation of $O(n)$(read big-O). The big-O notation describes the limiting behavior of a function or a constant when the parameters of the domain are changed. To illustrate, if $\phi \sim O(n)$ there exists a positive number $K$ such that $\phi < K \times n$ for all positive values of $n$. Now, the condition for equilibrium is that the forward rate of reaction must be equal to the backward rate of reaction. Thus one can write the following equations to characterize the equilibrium condition.
\begin{equation}\label{eq:eq_cond}
    k_1 c_A c_B = k_{-1}c_{AB}
\end{equation}
\begin{equation}\label{eq:eq_cond_nondim}
     \bar{c_A} \bar{c_B} = \bar{c_{AB}}
\end{equation}
Eq.~\ref{eq:eq_cond_nondim} is obtained from Eq~\ref{eq:eq_cond} by applying the scaling of the respective variables from Eq.~\ref{eq:scaling_B}. Observing Eq.~\ref{eq:nondim_AB} one can see that Eq.~\ref{eq:eq_cond_nondim} can be obtained from Eq.~\ref{eq:nondim_AB} if the time derivative($\partial c_{AB}/\partial t$) and the spatial derivative($\partial c_{AB}/\partial x^2$) are extremely small compared to the rest of the terms of Eq.~\ref{eq:nondim_AB}. The ratio of the diffusivity of the bound and the free enzyme, $\gamma$ is $O(1)$, a constant. Thus for the equilibrium condition to hold we need $\delta<< O(1)$, then only Eq.\ref{eq:nondim_AB} will yield Eq.\ref{eq:eq_cond_nondim}. \\
Also, $\beta$ is $O(1)$. For example, diffusivity of hexokinase reported previously is $7.2 \times 10^{-11} \hspace{1mm}m^2/s$\cite{Mohajerani2018} and the diffusivity of glucose is reported to be $6.6 \times 10^{-10} \hspace{1mm}m^2/s$\cite{ZIEGLER1987} which means that $\beta  =9.2$ which is $O(1)$. Therefore $\eta << O(1)$ or $\epsilon<<O(1)$ will also yield Eq.\ref{eq:eq_cond_nondim} from Eq.\ref{eq:nondim_B} and Eq.\ref{eq:nondim_A} respectively. Overall one needs $\epsilon,\delta,\eta$ to be much less than O(1) for the equilibrium condition to hold. This suggests that the equilibrium assumption cannot be used universally and even though most enzymes that have been used for chemotaxis experiments fulfil the aforementioned criterion we must be wary of them breaking down especially when the width of our channel narrows down significantly, or when the time taken to reach equilibrium is large. One can show using some crude assumptions(see SI) that the time for a system to reach equilibrium varies as $t_{eq} \sim O\big(\frac{1}{k_1c_{B0}+k_{-1}}\big)$. When the domain length narrows down considerably, the time scale of diffusion, $t_D \sim d^2/D_A$ will be comparable to the time taken to reach equilibrium($t_{eq} \sim O(t_D)$). In such a circumstance, one cannot assume instantaneous equilibrium in the system. 
\\
For this paper we have selected systems where the equilibrium will be valid. In such cases where the rate of the reaction is fast enough to make the time taken to reach equilibrium much smaller than the time taken for the protein to diffuse through the channel, one can simplify equations to the following:

\begin{equation}\label{eq:total_bal_A}
\frac{\partial c^{T}_A}{\partial t} = D_{A} \frac{\partial^2 c^{T}_A}{\partial x^2} + \alpha D_A \frac{\partial^2}{\partial x^2}\Big(\frac{c^{T}_A c_B}{K_d + c_B}\Big)
\end{equation}
\begin{equation}\label{eq:total_bal_B}
\frac{\partial c_B}{\partial t} = D_{B} \frac{\partial^2 c_B}{\partial x^2}
\end{equation}
\\
Here $c^{T}_A$ is the total protein concentration evaluated by adding Eq. ~\ref{eq:mass_bal_A} and ~\ref{eq:mass_bal_AB}. It is convenient to define the bound protein diffusivity as $D_{AB}  = D_A(1+\alpha)$\cite{Agudo-Canalejo2018}. A positive $\alpha$ means that the bound protein has a higher diffusivity than the free protein. When $\alpha$ is negative the diffusivity of the bound protein is lower than the free protein. The derivation of Eq. ~\ref{eq:total_bal_A} and ~\ref{eq:total_bal_B} is then completed by equilibrium assumptions $k_1 c_A c_B  = k_{-1}c_{AB}$ and $c_A + c_{AB}  = c^{T}_A$. We now non-dimensionlize the equations using the following parameters and obtain the final equations:
\begin{equation}\label{eq:scaling_eq}
    c^{T}_A \sim c^{T}_{A0} \hspace{1cm} c_B \sim c_{B0} \hspace{1cm} x \sim d \hspace{1cm} t \sim \frac{d^2}{D_B}
\end{equation}
\begin{equation}\label{eq:totalnondim_bal_A}
\frac{\partial \bar{c^{T}_A}}{\partial \bar{t}} = \frac{D_{A}}{D_B} \frac{\partial^2 \bar{c^{T}_A}}{\partial \bar{x}^2} + \alpha \frac{D_A}{D_B} \frac{\partial^2}{\partial \bar{x}^2}\Big(\frac{\bar{c^{T}_A} \bar{c_B}}{K_d/c_{B0} + \bar{c_B}}\Big)
\end{equation}
\begin{equation}\label{eq:total_balnondim_B}
\frac{\partial \bar{c_B}}{\partial \bar{t}} = \frac{\partial^2 \bar{c_B}}{\partial \bar{x}^2}
\end{equation}
Eq. ~\ref{eq:totalnondim_bal_A} and ~\ref{eq:total_balnondim_B} has been used for solving the concentration profiles in the 1D system shown in Figure ~\ref{fgr:1Dmicro}.

\paragraph{Enzymatic catalysis} An enzymatic reaction between an enzyme(E), substrate(S) and product(P) can be described as follows:
\begin{equation}\label{eqn:enzyme_main}
\ce{E + S <=>[k_1][k_{-1}] ES ->[k_2] E + P}
\end{equation}
The approach for analysing the reaction in Eq ~\ref{eqn:enzyme_main} remains same as that for the binding case except that the system never reaches equilibrium because of the irreversible catalytic step. It however can reach a pseudo-steady state where following relation hold:
\begin{equation}\label{eq:pssa}
    k_1 c_E c_S = (k_{-1}+k_2) c_{ES}
\end{equation}
The criterion for this relation to hold is to have large Damkolher numbers, $O(Da)>>1$. The detailed description of the validity of Eq.~\ref{eq:pssa} has been provided in the SI. Following the same approach of simplification and non-dimensionalization(the details of which can be found in the SI) as before we get the following final equations for the system involving enzymatic reaction:
\begin{equation}\label{eq:totalnondim_bal_E}
\frac{\partial \bar{c^{T}_E}}{\partial \bar{t}} = \frac{\partial^2 \bar{c^{T}_E}}{\partial \bar{x}^2} + \alpha \frac{\partial^2}{\partial \bar{x}^2}\Big(\frac{\bar{c^{T}_E} \bar{c_S}}{K_M/c_{S0} + \bar{c_S}}\Big)
\end{equation}
\begin{equation}\label{eq:total_balnondim_S}
\frac{\partial \bar{c_S}}{\partial \bar{t}} = \beta \frac{\partial^2 \bar{c_S}}{\partial \bar{x}^2} - k_2 \frac{c_{E0}}{c_{S0}}\frac{\bar{c^{T}_E} \bar{c_S}}{K_M/c_{S0} + \bar{c_S}}\frac{d^2}{D_E}
\end{equation}
Here $\bar{c^{T}_E},\bar{c_S}$ stand for the non-dimensionalized enzyme and substrate concentration respectively. $\alpha$ has the same definition as the previous section. $\beta = D_S/D_E$ is the ratio of diffusivity of the substrate to the diffusivity of the free enzyme. $K_M = \frac{k_{-1}+k_2}{k_1}$ is the Michaelis-Menten constant. Eq. ~\ref{eq:totalnondim_bal_E} and ~\ref{eq:total_balnondim_S} has been used for solving the concentration profiles in the 1D system shown in Figure ~\ref{fgr:1Dmicro}.
\section{Results}
\begin{figure}
  \includegraphics[scale =1]{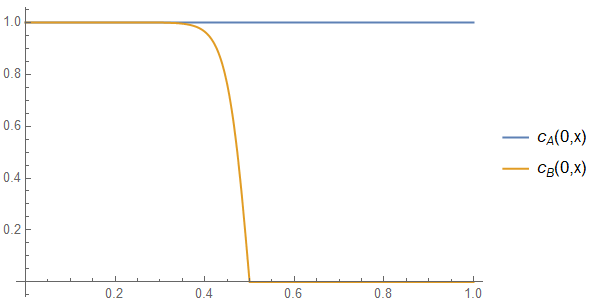}
  \caption{Non-dimensionalized initial condition for the protein/enzyme and the ligand/substrate in the domain.}
  \label{fgr:initial_cond}
\end{figure}

For both binding interaction and the enzymatic reaction, we have chosen a 1D geometry of width $d = 180\mu m$. We have placed at time $t=0$ a uniform concentration of protein/enzyme in the domain. The initial substrate concentration is however present at only one-half of the domain from $x = 0$  to $x = d/2$. The boundary condition at each end of the domain is a no-flux condition because the walls are impermeable. The initial conditions of the substrate and the enzyme in the domain can be seen in Figure~\ref{fgr:initial_cond}. Our goal is to see the time evolution of the concentration of total enzyme in this system for various values of $\alpha$. This analysis would tell us how $\alpha$ might affect chemotaxis in a time-dependent manner. 
\paragraph{Binding Interactions} We have summarized here the results for binding reaction described in Eq.~\ref{eqn:rxn_main}. We have selected hexokinase as our protein which reversibly binds with glucose to form an inactive complex in the absence of ATP in the system\cite{Wilkinson1981}. The forward and the backward reaction rates are $k_1 =2 \times 10^6 \hspace{1mm} M^{-1}s^{-1}$ and $k_{-1} = 60 \hspace{1mm} s^{-1}$ respectively\cite{Wilkinson1979}. Diffusivity of hexokinase is taken to be $7.2 \times 10^{-11} \hspace{1mm}m^2/s$. Diffusivity of glucose is taken to be 10 times faster than hexokinase. For our modeling, we set the initial ligand/protein concentration ratio at $10^4$.
\begin{figure}
    \centering
    \begin{subfigure}[t]{0.3\textwidth}
        \centering
        \includegraphics[width=\linewidth]{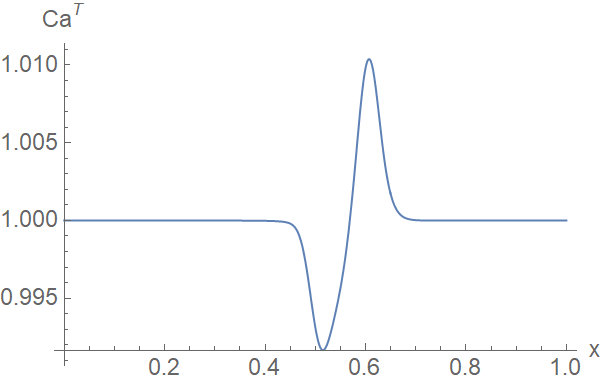} 
        \caption{$\alpha = 0.5$ t=0.001} \label{fig:alpha05t001}
    \end{subfigure}
    \hfill
    \begin{subfigure}[t]{0.3\textwidth}
        \centering
        \includegraphics[width=\linewidth]{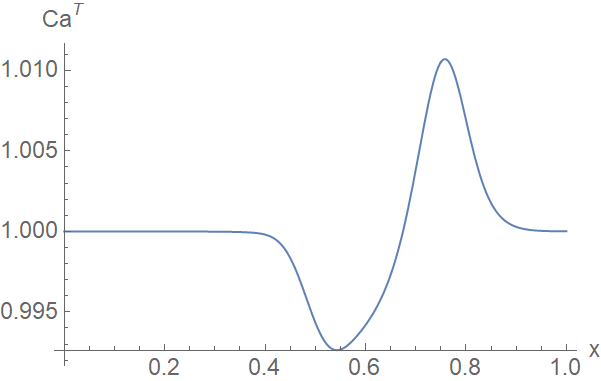} 
        \caption{$\alpha = 0.5$ t=0.005} \label{fig:alpha05t005}
    \end{subfigure}
    \hfill
    \begin{subfigure}[t]{0.3\textwidth}
        \centering
        \includegraphics[width=\linewidth]{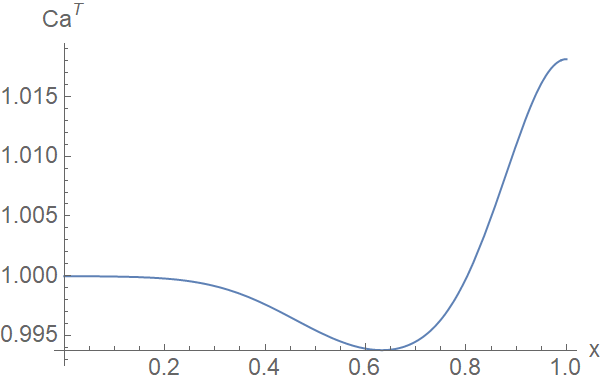} 
        \caption{$\alpha = 0.5$ t=0.06} \label{fig:alpha05t06}
    \end{subfigure}

    \vspace{1cm}
    \begin{subfigure}[t]{0.3\textwidth}
        \centering
        \includegraphics[width=\linewidth]{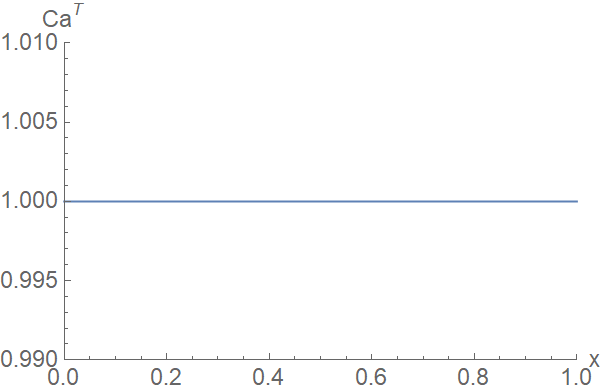} 
        \caption{$\alpha = 0$ t=0.001} \label{fig:alpha0t001}
    \end{subfigure}
    \hfill
    \begin{subfigure}[t]{0.3\textwidth}
        \centering
        \includegraphics[width=\linewidth]{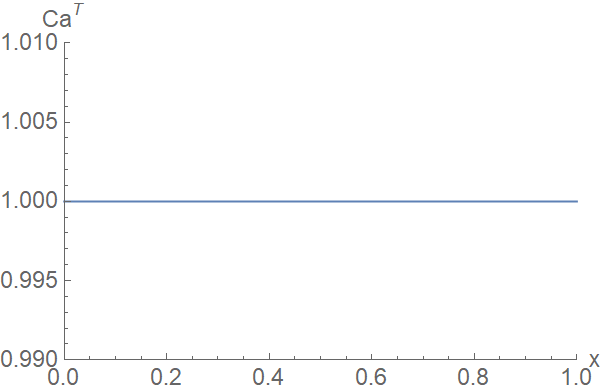} 
        \caption{$\alpha = 0$ t=0.005} \label{fig:alpha0t005}
    \end{subfigure}
    \hfill
    \begin{subfigure}[t]{0.3\textwidth}
        \centering
        \includegraphics[width=\linewidth]{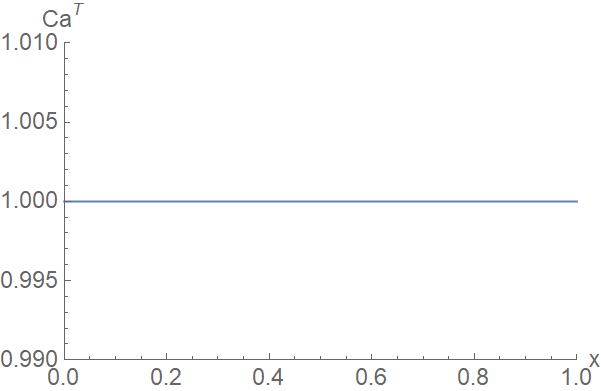} 
        \caption{$\alpha = 0$ t=0.06} \label{fig:alpha0t06}
    \end{subfigure}
    
    \vspace{1cm}
    \begin{subfigure}[t]{0.3\textwidth}
        \centering
        \includegraphics[width=\linewidth]{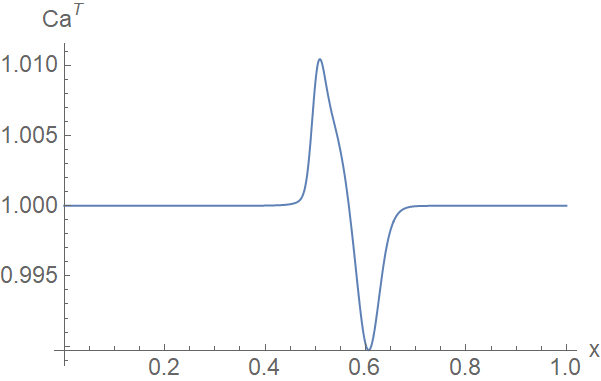} 
        \caption{$\alpha = -0.5$ t=0.001} \label{fig:alpham05t001}
    \end{subfigure}
    \hfill
    \begin{subfigure}[t]{0.3\textwidth}
        \centering
        \includegraphics[width=\linewidth]{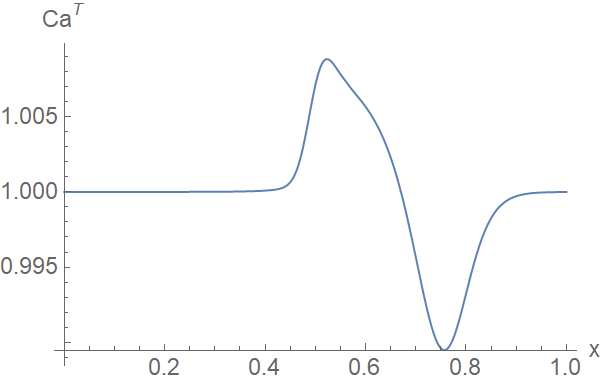} 
        \caption{$\alpha = -0.5$ t=0.005} \label{fig:alpham05t005}
    \end{subfigure}
    \hfill
    \begin{subfigure}[t]{0.3\textwidth}
        \centering
        \includegraphics[width=\linewidth]{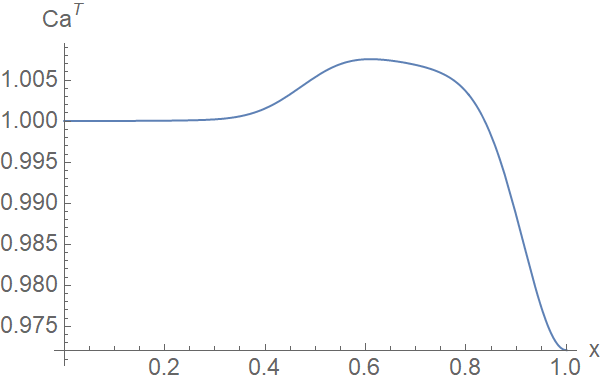} 
        \caption{$\alpha = -0.5$ t=0.06} \label{fig:alpham05t06}
    \end{subfigure}
    \caption{The concentration profile of total protein plotted against x-axis for various values of alpha. The concentration are non-dimensionalized with respect to the initial concentration of the enzyme. The time is also non-dimensionlized with respect to the diffusive time scale as shown in Eq.~\ref{eq:scaling_eq}}
    \label{fgr:binding_results}
\end{figure}
\\
Figure~\ref{fgr:binding_results} shows us the evolution of the concentration profile of total protein with time in the 1-D domain for three values of $\alpha$, one positive, one negative, and one zero. The time here is non-dimensionalized according to Eq.~\ref{eq:scaling_eq}. With the parameters for hexokinase, the diffusion time scale turns out to be $t_D \approx 450s$. That means non-dimensionalized time of $0.06$ corresponds to $t = 0.06 \times t_D \approx 27s$ which is typically the time used for chemotaxis experiments. The zero value for alpha shows that there will be no change in the concentration profile of the protein. Due to the conservation of protein mass and since the initial protein concentration is uniform everywhere the gradient of free protein and bound protein are equal in magnitude but act in opposite direction. We should note that the concentration of the free protein decreases in the direction of increasing substrate concentration since the concentration of the free protein will be higher at positions with less substrate. The opposite is true for the bound protein whose concentration increasing in the direction of increasing substrate concentration since bound protein will be found in higher concentration at positions where there is higher substrate concentration. The mass flux of total protein concentration is related to the concentration gradient of the free and bound species as follows:

\begin{equation}\label{eq:mass_flux}
    J_A^{T} = -D_A \frac{\partial c_A}{\partial x} -D_{AB} \frac{\partial c_{AB}}{\partial x}
\end{equation}
The zero value for alpha shows that there will be no change in the concentration profile of the protein since $D_A = D_{AB}$ and the two terms in Eq~\ref{eq:mass_flux} will cancel each other. For positive values of $\alpha$, i.e, when $D_{AB} > D_A$ the second term in Eq~\ref{eq:mass_flux} which is the mass-flux for bound protein, will be larger. The direction of this larger term is along decreasing concentration of substrate. Hence we will see that the total protein concentration will shift towards the direction of decreasing substrate concentration. This will result in negative chemotaxis. When the $\alpha$ values are negative we see the reverse effect happens. Now the gradient of the free protein has a greater contribution to the flux because of a higher diffusivity of the free protein. Since the gradient of the free protein is in the direction of increasing substrate concentration the protein will move towards the center of the channel. This is indeed what has been experimentally observed in microchannels\cite{Mohajerani2018}. From these results, it would seem that negative values of $\alpha$ which correspond to a less diffusive bound protein will lead to positive chemotaxis.
\paragraph{Enzymatic Catalysis} The same system of hexokinase has been analysed with an overall catalytic step that has a rate constant of $k_2 = 8000 s^{-1}$. For our modeling, we set the initial substrate/enzyme concentration ratio at $10^5$. We now solve for equations Eq.~\ref{eq:totalnondim_bal_E} and~\ref{eq:total_balnondim_S}.
\begin{figure}
    \centering
    \begin{subfigure}[t]{0.3\textwidth}
        \centering
        \includegraphics[width=\linewidth]{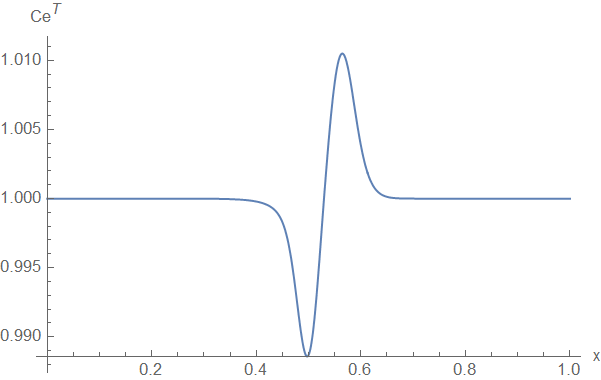} 
        \caption{$\alpha = 0.5$ t=0.0001} \label{fig:rxnalpha05t0001}
    \end{subfigure}
    \hfill
    \begin{subfigure}[t]{0.3\textwidth}
        \centering
        \includegraphics[width=\linewidth]{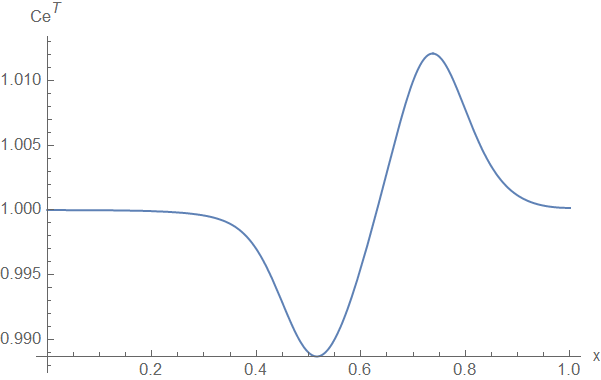} 
        \caption{$\alpha = 0.5$ t=0.001} \label{fig:rxnalpha05t001}
    \end{subfigure}
    \hfill
    \begin{subfigure}[t]{0.3\textwidth}
        \centering
        \includegraphics[width=\linewidth]{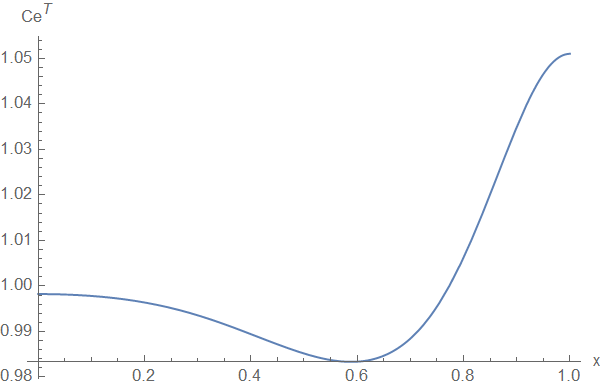} 
        \caption{$\alpha = 0.5$ t=0.01} \label{fig:rxnalpha05t01}
    \end{subfigure}

    \vspace{1cm}
    \begin{subfigure}[t]{0.3\textwidth}
        \centering
        \includegraphics[width=\linewidth]{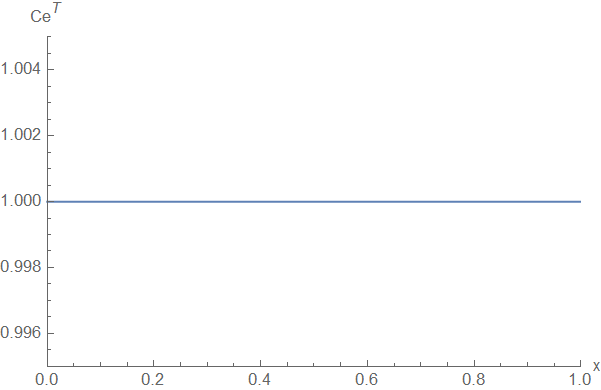} 
        \caption{$\alpha = 0$ t=0.0001} \label{fig:rxnalpha0t0001}
    \end{subfigure}
    \hfill
    \begin{subfigure}[t]{0.3\textwidth}
        \centering
        \includegraphics[width=\linewidth]{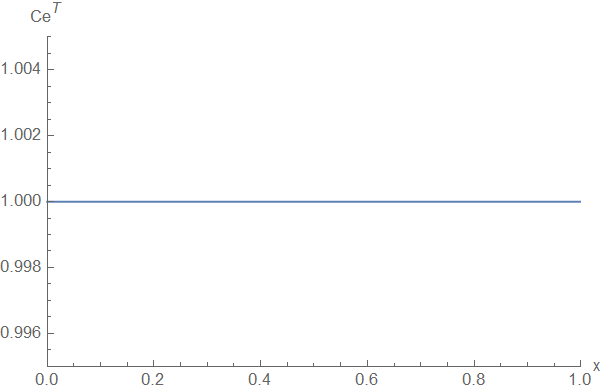} 
        \caption{$\alpha = 0$ t=0.001} \label{fig:rxnalpha0t001}
    \end{subfigure}
    \hfill
    \begin{subfigure}[t]{0.3\textwidth}
        \centering
        \includegraphics[width=\linewidth]{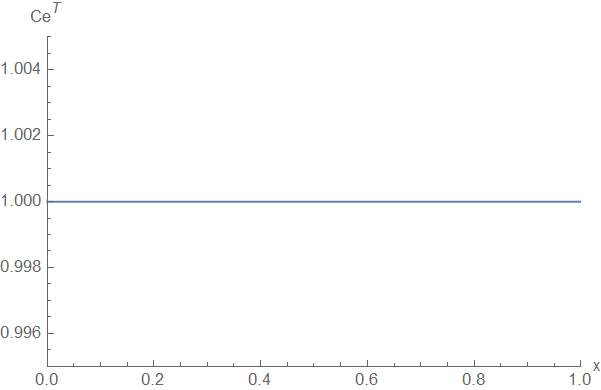} 
        \caption{$\alpha = 0$ t=0.01} \label{fig:rxnalpha0t01}
    \end{subfigure}
    
    \vspace{1cm}
    \begin{subfigure}[t]{0.3\textwidth}
        \centering
        \includegraphics[width=\linewidth]{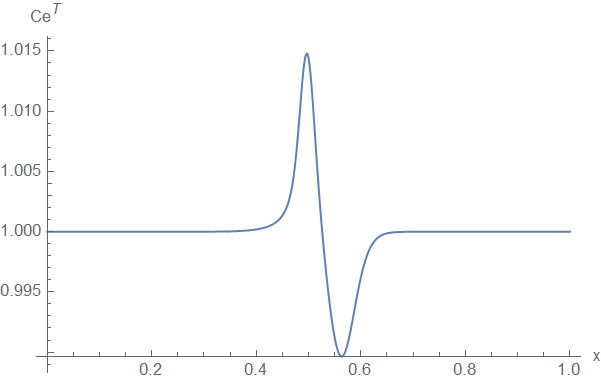} 
        \caption{$\alpha = -0.5$ t=0.0001} \label{fig:rxnalphan05t0001}
    \end{subfigure}
    \hfill
    \begin{subfigure}[t]{0.3\textwidth}
        \centering
        \includegraphics[width=\linewidth]{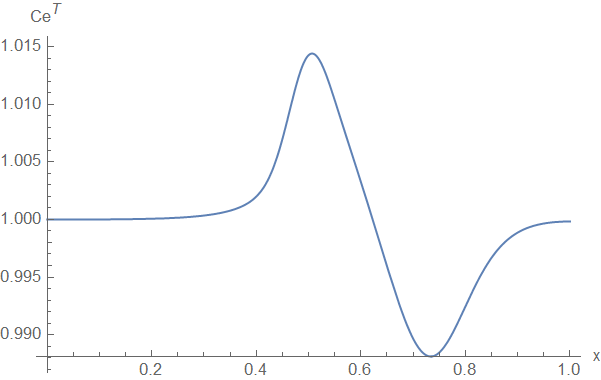} 
        \caption{$\alpha = -0.5$ t=0.001} \label{fig:rxnalphan05t001}
    \end{subfigure}
    \hfill
    \begin{subfigure}[t]{0.3\textwidth}
        \centering
        \includegraphics[width=\linewidth]{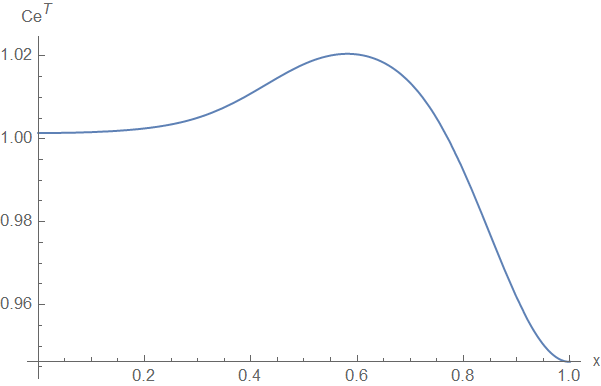} 
        \caption{$\alpha = -0.5$ t=0.01} \label{fig:rxnalphan05t01}
    \end{subfigure}
    \caption{The concentration profile of total enzyme plotted against x-axis for various values of alpha. The concentration are non-dimensionalized with respect to the initial concentration of the enzyme. The time is also non-dimensionlized with respect to the diffusive time scale as shown in SI}
    \label{fgr:reaction_results}
\end{figure}
Figure~\ref{fgr:reaction_results} shows the evolution of the total enzyme concentration with time for different values of $\alpha$. Similar to the binding interactions we see a reaction-front going away from the substrate for $\alpha>0$ and moving towards the substrate for $\alpha<0$. This would again suggest that for the focusing of enzymes to occur towards a higher concentration of substrate we will need $\alpha<0$. Compared to a simple binding interaction, the simulated concentration profile of the total enzyme develops faster because of catalysis.

\section{Discussion}

The primary deduction from the results shown above is that for positive chemotaxis to occur the diffusivity of the bound protein/enzyme must be lower than the free protein/enzyme. A higher diffusivity of the bound protein/enzyme will lead to the migration of the protein/enzyme away from the substrate, i.e. negative chemotaxis. Thus, it is critically important to measure the diffusivities of the free protein/enzyme and ligand/substrate-bound protein/enzyme.   
\paragraph{Reconciliation with theories on cross-diffusion}
Previous theories of enzyme chemotaxis have been based on cross-diffusion of an enzyme in response to a gradient of substrate\cite{XiZhao2018,Schurr2013,Mohajerani2018}. The overall approach for such a theory has been to use the total enzyme concentration at equilibrium in the system to characterize cross-diffusion. The focus is on the increased mass flux of free enzyme given the fact their equilibrium concentration will be lower than their original concentration because of the presence of the substrate. This paper through kinetic analysis explains the same idea without having to characterize a cross-diffusion phenomenon. We have also shown that the equilibrium assumption is not always valid and hence the basic formulation for any chemotactic system should always begin with Eq~\ref{eq:mass_bal_A}-~\ref{eq:mass_bal_B}. Starting from this point we can decide whether equilibrium or a pseudo-steady state assumption will be valid. The examples we have taken have high Damkolher numbers and have been selected because experimental observations are available for these enzymes. 
\paragraph{Why cannot we ignore diffusion of bound enzyme/protein species?}
We have seen through our analysis that bound protein/enzyme gradient exists in the direction of decreasing substrate concentration and hence diffuse in the direction opposite to the free enzyme. In a previous article \citeauthor{Mohajerani2018} explained chemotaxis by suggesting that the bound enzymes do not cross-diffuse\cite{Mohajerani2018}. This assumption corresponds to our case of $\alpha = -1$. That will be a limiting case where the bound protein/enzyme diffusivity will be exactly zero and both the self-diffusion and the cross-diffusion of the bound protein/enzyme are ignored. For other values of $\alpha$, we will need to include the bound protein/enzyme diffusivity. 
\section{Conclusion}

In this paper, we have shown how chemotaxis can be explained using kinetic analysis. We have formulated separate kinetic models for reactions with and without a turnover step, analyzed the validity of the equilibrium and the pseudo-steady-state assumptions in a mass transfer system, and discussed the conditions under which these assumptions are valid. For our modeling, we have assumed instantaneous achievement of equilibrium or pseud-steady state condition and hence we recognize that while our results are valid for the time scale of diffusion, the concentration profile at extremely small times needs to be explored. We have shown using numerical simulations how the difference in diffusivity between the free and bound enzyme manifests itself in a 1D domain. We see from the results that it is critically important to measure the diffusivity of the bound protein/enzyme and compare it with that of the free protein/enzyme. While a net increase in diffusion has been observed for enzymes during catalysis\cite{Muddana2010,Mohajerani2018,Zhang2019,Zhao2018,Xu2019,Riedel2015}, this increase in diffusion should not \textit{a priori} lead to the conclusion that the bound enzyme diffuses faster than the unbound counterpart. The enhancement in diffusion in active enzyme systems may also arise from thermodynamic driving forces associated with the binding and/or the turnover step\cite{Jee2020}. 
\\
We have also introduced in this area the practice of using finite difference methods rather than finite element methods. Finite difference methods being a differential scheme are faster than integral methods like FEM. At the same time for domains that are well described with cartesian systems like lines and rectangles, the finite difference method can increase its accuracy vastly by refining its grid.
\begin{acknowledgement}

The authors thank Prof Igor Aronson and Prof Ali Borhan for their insights in the way differential equations can be solved, and in understanding the scaling of variables for the governing equations. We acknowledge financial support by the Department of Energy, Office of Basic Energy Sciences (DOE-DE-SC0020964).

\end{acknowledgement}

\begin{suppinfo}

Kinetic analysis of enzyme catalysis, calculation of time taken to reach equilibrium, and simulation details. The information is available free of charge at https://pubs.acs.org/

\end{suppinfo}

\bibliography{achemso-demo}

\providecommand{\latin}[1]{#1}
\makeatletter
\providecommand{\doi}
  {\begingroup\let\do\@makeother\dospecials
  \catcode`\{=1 \catcode`\}=2 \doi@aux}
\providecommand{\doi@aux}[1]{\endgroup\texttt{#1}}
\makeatother
\providecommand*\mcitethebibliography{\thebibliography}
\csname @ifundefined\endcsname{endmcitethebibliography}
  {\let\endmcitethebibliography\endthebibliography}{}
\begin{mcitethebibliography}{29}
\providecommand*\natexlab[1]{#1}
\providecommand*\mciteSetBstSublistMode[1]{}
\providecommand*\mciteSetBstMaxWidthForm[2]{}
\providecommand*\mciteBstWouldAddEndPuncttrue
  {\def\EndOfBibitem{\unskip.}}
\providecommand*\mciteBstWouldAddEndPunctfalse
  {\let\EndOfBibitem\relax}
\providecommand*\mciteSetBstMidEndSepPunct[3]{}
\providecommand*\mciteSetBstSublistLabelBeginEnd[3]{}
\providecommand*\EndOfBibitem{}
\mciteSetBstSublistMode{f}
\mciteSetBstMaxWidthForm{subitem}{(\alph{mcitesubitemcount})}
\mciteSetBstSublistLabelBeginEnd
  {\mcitemaxwidthsubitemform\space}
  {\relax}
  {\relax}

\bibitem[Berg(2003)]{Berg1975}
Berg,~H.~C. Chemotaxis in Bacteria. \emph{Annu. Rev. Biophys. Bioeng.}
  \textbf{2003}, \emph{125}, 1128--1129\relax
\mciteBstWouldAddEndPuncttrue
\mciteSetBstMidEndSepPunct{\mcitedefaultmidpunct}
{\mcitedefaultendpunct}{\mcitedefaultseppunct}\relax
\EndOfBibitem
\bibitem[Sengupta \latin{et~al.}(2013)Sengupta, Dey, Muddana, Tabouillot,
  Ibele, Butler, and Sen]{Sengupta2013}
Sengupta,~S.; Dey,~K.~K.; Muddana,~H.~S.; Tabouillot,~T.; Ibele,~M.~E.;
  Butler,~P.~J.; Sen,~A. {Enzyme molecules as nanomotors}. \emph{J. Am. Chem.
  Soc.} \textbf{2013}, \emph{135}, 1406--1414\relax
\mciteBstWouldAddEndPuncttrue
\mciteSetBstMidEndSepPunct{\mcitedefaultmidpunct}
{\mcitedefaultendpunct}{\mcitedefaultseppunct}\relax
\EndOfBibitem
\bibitem[Dey \latin{et~al.}(2014)Dey, Das, Poyton, Sengupta, Butler, Cremer,
  and Sen]{Dey2014}
Dey,~K.~K.; Das,~S.; Poyton,~M.~F.; Sengupta,~S.; Butler,~P.~J.; Cremer,~P.~S.;
  Sen,~A. {Chemotactic separation of enzymes}. \emph{ACS Nano} \textbf{2014},
  \emph{8}, 11941--11949\relax
\mciteBstWouldAddEndPuncttrue
\mciteSetBstMidEndSepPunct{\mcitedefaultmidpunct}
{\mcitedefaultendpunct}{\mcitedefaultseppunct}\relax
\EndOfBibitem
\bibitem[Zhao \latin{et~al.}(2018)Zhao, Palacci, Yadav, Spiering, Gilson,
  Butler, Hess, Benkovic, and Sen]{XiZhao2018}
Zhao,~X.; Palacci,~H.; Yadav,~V.; Spiering,~M.~M.; Gilson,~M.~K.;
  Butler,~P.~J.; Hess,~H.; Benkovic,~S.~J.; Sen,~A. {Substrate-driven
  chemotactic assembly in an enzyme cascade}. \emph{Nat. Chem.} \textbf{2018},
  \emph{10}, 311--317\relax
\mciteBstWouldAddEndPuncttrue
\mciteSetBstMidEndSepPunct{\mcitedefaultmidpunct}
{\mcitedefaultendpunct}{\mcitedefaultseppunct}\relax
\EndOfBibitem
\bibitem[Jee \latin{et~al.}(2018)Jee, Dutta, Cho, Tlusty, and Granick]{Jee2018}
Jee,~A.~Y.; Dutta,~S.; Cho,~Y.~K.; Tlusty,~T.; Granick,~S. {Enzyme leaps fuel
  antichemotaxis}. \emph{Proc. Natl. Acad. Sci. U. S. A.} \textbf{2018},
  \emph{115}, 14--18\relax
\mciteBstWouldAddEndPuncttrue
\mciteSetBstMidEndSepPunct{\mcitedefaultmidpunct}
{\mcitedefaultendpunct}{\mcitedefaultseppunct}\relax
\EndOfBibitem
\bibitem[Somasundar \latin{et~al.}(2019)Somasundar, Ghosh, Mohajerani,
  Massenburg, Yang, Cremer, Velegol, and Sen]{Somasundar2019}
Somasundar,~A.; Ghosh,~S.; Mohajerani,~F.; Massenburg,~L.~N.; Yang,~T.;
  Cremer,~P.~S.; Velegol,~D.; Sen,~A. Positive and negative chemotaxis of
  enzyme-coated liposome motors. \emph{Nat. Nanotechnol.} \textbf{2019},
  \emph{14}, 1129--1134\relax
\mciteBstWouldAddEndPuncttrue
\mciteSetBstMidEndSepPunct{\mcitedefaultmidpunct}
{\mcitedefaultendpunct}{\mcitedefaultseppunct}\relax
\EndOfBibitem
\bibitem[Dey \latin{et~al.}(2015)Dey, Zhao, Tansi, M{\'{e}}ndez-Ortiz,
  C{\'{o}}rdova-Figueroa, Golestanian, and Sen]{Dey2015}
Dey,~K.~K.; Zhao,~X.; Tansi,~B.~M.; M{\'{e}}ndez-Ortiz,~W.~J.;
  C{\'{o}}rdova-Figueroa,~U.~M.; Golestanian,~R.; Sen,~A. {Micromotors Powered
  by Enzyme Catalysis}. \emph{Nano Lett.} \textbf{2015}, \emph{15},
  8311--8315\relax
\mciteBstWouldAddEndPuncttrue
\mciteSetBstMidEndSepPunct{\mcitedefaultmidpunct}
{\mcitedefaultendpunct}{\mcitedefaultseppunct}\relax
\EndOfBibitem
\bibitem[Ji \latin{et~al.}(2019)Ji, Lin, Wu, Wu, Gao, and He]{Ji2019}
Ji,~Y.; Lin,~X.; Wu,~Z.; Wu,~Y.; Gao,~W.; He,~Q. {Macroscale Chemotaxis from a
  Swarm of Bacteria-Mimicking Nanoswimmers}. \emph{Angew. Chemie - Int. Ed.}
  \textbf{2019}, \emph{58}, 12200--12205\relax
\mciteBstWouldAddEndPuncttrue
\mciteSetBstMidEndSepPunct{\mcitedefaultmidpunct}
{\mcitedefaultendpunct}{\mcitedefaultseppunct}\relax
\EndOfBibitem
\bibitem[Joseph \latin{et~al.}(2017)Joseph, Contini, Cecchin, Nyberg,
  Ruiz-Perez, Gaitzsch, Fullstone, Tian, Azizi, Preston, Volpe, and
  Battaglia]{Joseph2017}
Joseph,~A.; Contini,~C.; Cecchin,~D.; Nyberg,~S.; Ruiz-Perez,~L.; Gaitzsch,~J.;
  Fullstone,~G.; Tian,~X.; Azizi,~J.; Preston,~J.; Volpe,~G.; Battaglia,~G.
  {Chemotactic synthetic vesicles: Design and applications in blood-brain
  barrier crossing}. \emph{Sci. Adv.} \textbf{2017}, \emph{3}\relax
\mciteBstWouldAddEndPuncttrue
\mciteSetBstMidEndSepPunct{\mcitedefaultmidpunct}
{\mcitedefaultendpunct}{\mcitedefaultseppunct}\relax
\EndOfBibitem
\bibitem[Wang \latin{et~al.}(2020)Wang, Toebes, Plachokova, Liu, Deng, Jansen,
  Yang, and Wilson]{Wang2020}
Wang,~J.; Toebes,~B.~J.; Plachokova,~A.~S.; Liu,~Q.; Deng,~D.; Jansen,~J.~A.;
  Yang,~F.; Wilson,~D.~A. {Self-Propelled PLGA Micromotor with Chemotactic
  Response to Inflammation}. \emph{Adv. Healthc. Mater.} \textbf{2020},
  \emph{9}, 1--8\relax
\mciteBstWouldAddEndPuncttrue
\mciteSetBstMidEndSepPunct{\mcitedefaultmidpunct}
{\mcitedefaultendpunct}{\mcitedefaultseppunct}\relax
\EndOfBibitem
\bibitem[Feng and Gilson(2020)Feng, and Gilson]{Feng2020}
Feng,~M.; Gilson,~M.~K. {Enhanced Diffusion and Chemotaxis of Enzymes}.
  \emph{Annu. Rev. Biophys.} \textbf{2020}, \emph{49}, 87--105\relax
\mciteBstWouldAddEndPuncttrue
\mciteSetBstMidEndSepPunct{\mcitedefaultmidpunct}
{\mcitedefaultendpunct}{\mcitedefaultseppunct}\relax
\EndOfBibitem
\bibitem[Agudo-Canalejo \latin{et~al.}(2018)Agudo-Canalejo, Illien, and
  Golestanian]{Agudo-Canalejo2018}
Agudo-Canalejo,~J.; Illien,~P.; Golestanian,~R. {Phoresis and Enhanced
  Diffusion Compete in Enzyme Chemotaxis}. \emph{Nano Lett.} \textbf{2018},
  \emph{18}, 2711--2717\relax
\mciteBstWouldAddEndPuncttrue
\mciteSetBstMidEndSepPunct{\mcitedefaultmidpunct}
{\mcitedefaultendpunct}{\mcitedefaultseppunct}\relax
\EndOfBibitem
\bibitem[Schurr \latin{et~al.}(2013)Schurr, Fujimoto, Huynh, and
  Chiu]{Schurr2013}
Schurr,~J.~M.; Fujimoto,~B.~S.; Huynh,~L.; Chiu,~D.~T. {A theory of
  macromolecular chemotaxis}. \emph{J. Phys. Chem. B} \textbf{2013},
  \emph{117}, 7626--7652\relax
\mciteBstWouldAddEndPuncttrue
\mciteSetBstMidEndSepPunct{\mcitedefaultmidpunct}
{\mcitedefaultendpunct}{\mcitedefaultseppunct}\relax
\EndOfBibitem
\bibitem[Zhao \latin{et~al.}(2018)Zhao, Gentile, Mohajerani, and Sen]{Zhao2018}
Zhao,~X.; Gentile,~K.; Mohajerani,~F.; Sen,~A. {Powering Motion with Enzymes}.
  \emph{Acc. Chem. Res.} \textbf{2018}, \emph{51}, 2373--2381\relax
\mciteBstWouldAddEndPuncttrue
\mciteSetBstMidEndSepPunct{\mcitedefaultmidpunct}
{\mcitedefaultendpunct}{\mcitedefaultseppunct}\relax
\EndOfBibitem
\bibitem[Mohajerani \latin{et~al.}(2018)Mohajerani, Zhao, Somasundar, Velegol,
  and Sen]{Mohajerani2018}
Mohajerani,~F.; Zhao,~X.; Somasundar,~A.; Velegol,~D.; Sen,~A. {A Theory of
  Enzyme Chemotaxis: From Experiments to Modeling}. \emph{Biochemistry}
  \textbf{2018}, \emph{57}, 6256--6263\relax
\mciteBstWouldAddEndPuncttrue
\mciteSetBstMidEndSepPunct{\mcitedefaultmidpunct}
{\mcitedefaultendpunct}{\mcitedefaultseppunct}\relax
\EndOfBibitem
\bibitem[Blazek(2015)]{Blazek2015}
Blazek,~J. \emph{Computational Fluid Dynamics: Principles and Applications},
  3rd ed.; Elsevier Science \& Technology: Waltham, MA, 2015\relax
\mciteBstWouldAddEndPuncttrue
\mciteSetBstMidEndSepPunct{\mcitedefaultmidpunct}
{\mcitedefaultendpunct}{\mcitedefaultseppunct}\relax
\EndOfBibitem
\bibitem[Lin and Butcher(2006)Lin, and Butcher]{Lin2006}
Lin,~F.; Butcher,~E.~C. {T cell chemotaxis in a simple microfluidic device}.
  \emph{Lab Chip} \textbf{2006}, \emph{6}, 1462--1469\relax
\mciteBstWouldAddEndPuncttrue
\mciteSetBstMidEndSepPunct{\mcitedefaultmidpunct}
{\mcitedefaultendpunct}{\mcitedefaultseppunct}\relax
\EndOfBibitem
\bibitem[Diao \latin{et~al.}(2006)Diao, Young, Kim, Fogarty, Heilman, Zhou,
  Shuler, Wu, and DeLisa]{Diao2006}
Diao,~J.; Young,~L.; Kim,~S.; Fogarty,~E.~A.; Heilman,~S.~M.; Zhou,~P.;
  Shuler,~M.~L.; Wu,~M.; DeLisa,~M.~P. {A three-channel microfluidic device for
  generating static linear gradients and its application to the quantitative
  analysis of bacterial chemotaxis}. \emph{Lab Chip} \textbf{2006}, \emph{6},
  381--388\relax
\mciteBstWouldAddEndPuncttrue
\mciteSetBstMidEndSepPunct{\mcitedefaultmidpunct}
{\mcitedefaultendpunct}{\mcitedefaultseppunct}\relax
\EndOfBibitem
\bibitem[Fogler(2006)]{Fogler2006}
Fogler,~S.~H. \emph{Elements of Chemical Reaction Engineering}, 4th ed.;
  Pearson: Essex, United Kingdom, 2006\relax
\mciteBstWouldAddEndPuncttrue
\mciteSetBstMidEndSepPunct{\mcitedefaultmidpunct}
{\mcitedefaultendpunct}{\mcitedefaultseppunct}\relax
\EndOfBibitem
\bibitem[Deen(2012)]{Deen2011}
Deen,~W.~M. \emph{Analysis of Transport Phenomena}, 2nd ed.; Oxford University
  Press: New York, 2012\relax
\mciteBstWouldAddEndPuncttrue
\mciteSetBstMidEndSepPunct{\mcitedefaultmidpunct}
{\mcitedefaultendpunct}{\mcitedefaultseppunct}\relax
\EndOfBibitem
\bibitem[Ziegler \latin{et~al.}(1987)Ziegler, Benado, and Rizvi]{ZIEGLER1987}
Ziegler,~G.~R.; Benado,~A.~L.; Rizvi,~S.~S. {Determination of Mass Diffusivity
  of Simple Sugars in Water by the Rotating Disk Method}. \emph{J. Food Sci.}
  \textbf{1987}, \emph{52}, 501--502\relax
\mciteBstWouldAddEndPuncttrue
\mciteSetBstMidEndSepPunct{\mcitedefaultmidpunct}
{\mcitedefaultendpunct}{\mcitedefaultseppunct}\relax
\EndOfBibitem
\bibitem[Wilkinson and Rose(1981)Wilkinson, and Rose]{Wilkinson1981}
Wilkinson,~K.~D.; Rose,~I.~A. {Study of crystalline hexokinase-glucose
  complexes by isotope trapping.} \emph{J. Biol. Chem.} \textbf{1981},
  \emph{256}, 9890--9894\relax
\mciteBstWouldAddEndPuncttrue
\mciteSetBstMidEndSepPunct{\mcitedefaultmidpunct}
{\mcitedefaultendpunct}{\mcitedefaultseppunct}\relax
\EndOfBibitem
\bibitem[Wilkinson and Rose(1979)Wilkinson, and Rose]{Wilkinson1979}
Wilkinson,~K.; Rose,~I. {Isotope trapping studies of yeast hexokinase during
  steady state catalysis}. \emph{J. Biol. Chem.} \textbf{1979}, \emph{254},
  12567--12572\relax
\mciteBstWouldAddEndPuncttrue
\mciteSetBstMidEndSepPunct{\mcitedefaultmidpunct}
{\mcitedefaultendpunct}{\mcitedefaultseppunct}\relax
\EndOfBibitem
\bibitem[Muddana \latin{et~al.}(2010)Muddana, Sengupta, Mallouk, Sen, and
  Butler]{Muddana2010}
Muddana,~H.~S.; Sengupta,~S.; Mallouk,~T.~E.; Sen,~A.; Butler,~P.~J. {Substrate
  catalysis enhances single-enzyme diffusion}. \emph{J. Am. Chem. Soc.}
  \textbf{2010}, \emph{132}, 2110--2111\relax
\mciteBstWouldAddEndPuncttrue
\mciteSetBstMidEndSepPunct{\mcitedefaultmidpunct}
{\mcitedefaultendpunct}{\mcitedefaultseppunct}\relax
\EndOfBibitem
\bibitem[Zhang and Hess(2019)Zhang, and Hess]{Zhang2019}
Zhang,~Y.; Hess,~H. {Enhanced Diffusion of Catalytically Active Enzymes}.
  \emph{ACS Cent. Sci.} \textbf{2019}, \emph{5}, 939--948\relax
\mciteBstWouldAddEndPuncttrue
\mciteSetBstMidEndSepPunct{\mcitedefaultmidpunct}
{\mcitedefaultendpunct}{\mcitedefaultseppunct}\relax
\EndOfBibitem
\bibitem[Xu \latin{et~al.}(2019)Xu, Ross, Valdez, and Sen]{Xu2019}
Xu,~M.; Ross,~J.~L.; Valdez,~L.; Sen,~A. {Direct Single Molecule Imaging of
  Enhanced Enzyme Diffusion}. \emph{Phys. Rev. Lett.} \textbf{2019},
  \emph{123}, 1--5\relax
\mciteBstWouldAddEndPuncttrue
\mciteSetBstMidEndSepPunct{\mcitedefaultmidpunct}
{\mcitedefaultendpunct}{\mcitedefaultseppunct}\relax
\EndOfBibitem
\bibitem[Riedel \latin{et~al.}(2015)Riedel, Gabizon, Wilson, Hamadani,
  Tsekouras, Marqusee, Press{\'{e}}, and Bustamante]{Riedel2015}
Riedel,~C.; Gabizon,~R.; Wilson,~C.~A.; Hamadani,~K.; Tsekouras,~K.;
  Marqusee,~S.; Press{\'{e}},~S.; Bustamante,~C. {The heat released during
  catalytic turnover enhances the diffusion of an enzyme}. \emph{Nature}
  \textbf{2015}, \emph{517}, 227--230\relax
\mciteBstWouldAddEndPuncttrue
\mciteSetBstMidEndSepPunct{\mcitedefaultmidpunct}
{\mcitedefaultendpunct}{\mcitedefaultseppunct}\relax
\EndOfBibitem
\bibitem[Jee \latin{et~al.}(2020)Jee, Tlusty, and Granick]{Jee2020}
Jee,~A.~Y.; Tlusty,~T.; Granick,~S. {Master curve of boosted diffusion for 10
  catalytic enzymes}. \emph{Proc. Natl. Acad. Sci. U. S. A.} \textbf{2020},
  \emph{117}, 29435--29441\relax
\mciteBstWouldAddEndPuncttrue
\mciteSetBstMidEndSepPunct{\mcitedefaultmidpunct}
{\mcitedefaultendpunct}{\mcitedefaultseppunct}\relax
\EndOfBibitem
\end{mcitethebibliography}

\end{document}